\documentclass[aps,prl,oneside,twocolumn]{revtex4-1}
\usepackage{amsmath}
\usepackage{esint}
\usepackage{amssymb}
\usepackage{graphicx}
\usepackage{mathtools}

\begin{document}

\draft \title{Polynomial potentials determined from the energy spectrum and transition dipole moments that give the largest hyperpolarizabilities}
\author{Nathan J. Dawson}
\affiliation{Department of physics, College of The Bahamas, Nassau, Bahamas}
\email{dawsphys@hotmail.com}
\affiliation{Department of physics and astronomy, Washington State University, Pullman, WA, 99164, USA}
\author{Mark G. Kuzyk}
\affiliation{Department of physics and astronomy, Washington State University, Pullman, WA, 99164, USA}
\date{\today}

\begin{abstract}
We attempt to get a polynomial solution to the inverse problem, that is, to determine the form of the mechanical Hamiltonian when given the energy spectrum and transition dipole moment matrix. Our approach is to determine the potential in the form of a polynomial by finding an approximate solution to the inverse problem, then to determine the hyperpolarizability for that system's Hamiltonian. We find that the largest hyperpolarizabilities approach the apparent limit of previous potential optimization studies, but we do not find real potentials for the parameter values necessary to exceed this apparent limit.  We also explore half potentials with positive exponent, which cannot be expressed as a polynomial except for integer powers.  This yields a simple closed potential with only one parameter that scans nearly the full range of the intrinsic hyperpolarizability.  The limiting case of vanishing exponent yields the largest intrinsic hyperpolarizability.
\end{abstract}

\maketitle

\section{Introduction}

The core component photonics industry was estimated to be \$182 billion USD near the end of 2015,\cite{unknown15.oct} where many of these technologies use nonlinear optical materials. The fundamental limit of nonlinear optical coefficients \cite{kuzyk00.01} derived from the Thomas-Reiche-Kuhn (TRK) sum rules \cite{thoma25.01,reich25.01,kuhn25.01} using a three-level ansatz restrict the intrinsic nonlinear response of molecules to a maximum value based on the energy spacing of the two lowest states and the number of electrons. Since the discovery of these limits, several studies on the shape of quantum wells have been undertaken in an attempt to find quantum potentials that achieve this limit.\cite{ather12.01} Thus far, such attempts using optimization techniques for one- \cite{zhou06.01,zhou07.02} and two-dimensional \cite{shafe12.01,lytel15.01,lytel15.02} systems have been unable to achieve the limit, where there appears to be a gap between the largest hyperpolarizability from optimized potential calculations and the fundamental limit.\cite{kuzyk13.01} In addition, there also appears to be a gap between the largest known intrinsic hyperpolarizabilities via measurements of real molecules and the largest value obtained by optimizing mathematical potentials.\cite{Tripa04.01,perez07.01,perez07.02,kuzyk13.01} Recently, new approaches to the design of real molecules with larger nonlinear optical coefficients have been explored including systematic searches using multipolar charge-density analysis from crystallographic data, \cite{cole02.01,cole03.01,higgi12.01} extended conjugation of donor-acceptor molecules,\cite{May07.01,stefk13.01} twisted donor acceptor molecules,\cite{Kang05.01,Kang07.01,brown08.01,shi15.02} and plasmonic clusters.\cite{knopp12.01,knopp15.01}

The fundamental limit is derived from the Hamiltonian appearing in the Schr\"{o}dinger equation using the commutative relationship between the canonical position and momentum operators. Therefore, the position-dependent potential does not affect the fundamental limit, making it a fully general result that must be obeyed by all quantum systems so described. The observed gap between the largest hyperpolarizabilities calculated from a potential and the fundamental limit is of concern.  One explanation for the difference is that the parameters used to calculate the limit from the TRK sum rules may be unphysical.  This notion has led researchers to recently consider more exotic Hamiltonians in search of the limit.\cite{watki12.01,shafe13.01,burke13.01,dawson15.01}

In an attempt to understand the gap between the previously-reported largest hyperpolarizability from optimized scale-invariant potentials and the fundamental limit, we search for potentials that might yield the largest response.  Our approach, which is independent of the others tried to date, seeks to find an approximate solution to the inverse problem of finding the potential that gives the largest values from the energy spectrum and transition dipole matrix. This is related to the famous problem given by Kac. \cite{kac66.01}  The ideal energy spacing to get hyperpolarizabilities at the limit can be determined from Monte Carlo techniques;\cite{kuzyk08.01} but, there is no reliable way to get the underlying Hamiltonian. We assume that the potential can be approximated by a polynomial, which is a truncation of the full power series that describes the potential.  Such polynomials, if enough terms are included, can approximate most well-behaved functions, where the range of validity and accuracy of the polynomial representation of a function increases as the number of terms is increased.  In contrast, a real Laurent series can also represent scale-invariant, singular potentials, and may be an approach for future consideration.

The current approach is tested by comparing the transition dipole moments and energy spectrum calculated from the approximate polynomial potential to the initial transition dipole moments and energy spectrum of a multilevel system.  We also test a locus of energy spectra and transition dipole moments in the neighborhood of the resultant parameters corresponding to the limiting case to help understand the apparent gap between the hyperpolarizability calculated from previously studied classes of potentials and the limit calculated from the TRK sum rules under the constraints of the three level ansatz.

Using this approach, we find the same apparent limit as was found in other past studies that can be found in the literature, giving one more piece of confirming evidence that the lower limit may be the true one.\cite{kuzyk13.01}

\section{Theory}

We study systems that can be described by a one-dimensional, single-particle Hamiltonian of the form
\begin{equation}
H = \frac{p^2}{2 m} + V\left(x\right) ,
\label{eq:hamilt}
\end{equation}
where $p$ is the momentum operator, $m$ is the mass, and $V\left(x\right)$ is the spatially dependent potential. We consider those cases in which the potential may be written as a power series in the position of the form
\begin{equation}
V\left(x\right) = \displaystyle \sum_{q = 0}^{\infty} a_q \left(x-c\right)^q ,
\label{eq:potpower}
\end{equation}
where the $a_q$ are the series coefficients and $c$ is a constant. The time-independent Schr\"{o}dinger equation is then given by
\begin{equation}
\left[\frac{p^2}{2 m} + \displaystyle \sum_{q = 0}^{\infty} a_q \left(x-c\right)^q \right] \left|n\right\rangle = E_n \left|n\right\rangle ,
\label{eq:schrod}
\end{equation}
where $E_n$ is the energy of eigenstate $\left| n \right\rangle$.

Taking the inner product with $\left\langle \ell \right|$ and using closure,
\begin{equation}
\mathbb{1} = \displaystyle \sum_{i=0}^{\infty}  \left| i \right\rangle \left\langle i \right|,
\label{eq:closure}
\end{equation}
we can rewrite Eq. \ref{eq:schrod} as the matrix equation
\begin{align}
&\sum_{q=0}^{\infty}\,\, \sum_{i_1,\ldots,i_q=0}^{\infty} \hspace{-0.3cm} a_q \left(x_{\ell i_1}-c\, \delta_{\ell i_1}\right) \left(x_{i_1 i_2} -c\, \delta_{i_1 i_2} \right) \ldots \nonumber \\
&\ldots \left(x_{i_q n}-c\, \delta_{i_q n} \right) + \frac{1}{2m}\displaystyle \sum_{i=0}^{\infty} p_{\ell i} p_{i n}  = \delta_{\ell n} E_n .
\label{eq:fullschrod}
\end{align}
We have used shorthand notation in Eq. \ref{eq:fullschrod} where ${\cal A}_{\ell n} = \left\langle \ell \right| {\cal A} \left| n \right\rangle$ for the operator ${\cal A}$.

Taking the matrix elements of the commutator $\left \langle l \right| \left[x,H\right] \left| n \right\rangle$, we can express momentum matrix elements as
\begin{equation}
p_{n \ell} = -i\frac{m}{\hbar} E_{n \ell} x_{n \ell} ,
\label{eq:momentumME}
\end{equation}
where $E_{n \ell} = E_n - E_\ell$. Substituting Eq. \ref{eq:momentumME} into Eq. \ref{eq:fullschrod} and setting $c = x_{00}$ gives
\begin{align}
&-\frac{m}{2\hbar^2} \displaystyle \sum_{i=0}^{\infty} E_{\ell i} E_{i n} x_{\ell i} x_{i n} + \sum_{q=0}^{\infty}\,\, \sum_{i_1,\ldots,i_q=0}^{\infty} \hspace{-.3cm} a_q \overline{x}_{\ell i_1} \overline{x}_{i_1 i_2} \ldots \overline{x}_{i_q n} \nonumber \\
&= \delta_{\ell n} E_n ,
\label{eq:schrodME}
\end{align}
where $\overline{x}_{ij} = x_{ij} - x_{00}\,\delta_{ij}$. Defining
\begin{equation}
B_{n \ell} \left(q\right) = \hspace{-0.2cm} \sum_{i_1,\ldots,i_q=0}^{\infty} \hspace{-0.3cm}\overline{x}_{\ell i_1} \overline{x}_{i_1 i_2} \ldots \overline{x}_{i_q n}
\label{eq:Bq}
\end{equation}
and
\begin{equation}
C_{n \ell} = \frac{m}{2\hbar^2} \displaystyle \sum_{i=0}^{\infty} E_{\ell i} E_{i n} x_{\ell i} x_{i n} + \delta_{n \ell} E_n ,
\label{eq:Cq}
\end{equation}
Eq. \ref{eq:schrodME} can be re-expressed in the simple form
\begin{equation}
\displaystyle \sum_{q=0}^{\infty} a_{q} B_{n \ell} \left(q\right) = C_{n \ell} .
\label{eq:ABCq}
\end{equation}

The approach is to find a set of coefficients $a_q$ in Eq. \ref{eq:ABCq} for a truncated $N$-state model in terms of only the transition dipole moments and energies.  Because the transition dipole moment matrix is Hermitian, the matrix must be symmetric when the matrix elements are real, \textit{i}.\textit{e}. $x_{ij} = x_{ji}$, which limits Eq. \ref{eq:ABCq} to at most $M$ linearly independent equations, where $M$ is given by the triangular number
\begin{equation}
M = \displaystyle \sum_{n=1}^{N} n = \frac{1}{2}\left(N+1\right) N.
\label{eq:indequations}
\end{equation}

\section{Method}

The matrix equation given by Eq. \ref{eq:ABCq} may be written in the form
\begin{equation}
B^\prime a = c \, .
\label{eq:explicitBpac}
\end{equation}
Setting $k = N-1$ as the largest excited state, the $B^\prime$ matrix is given as
\begin{equation}
\left( \begin{array}{cccccc}
B_{00}\left(0\right) & B_{00}\left(1\right) & B_{00}\left(2\right) & B_{00}\left(3\right) & \cdots & B_{00}\left(M-1\right) \\
B_{01}\left(0\right) & B_{01}\left(1\right) & B_{01}\left(2\right) & B_{01}\left(3\right) & \cdots & B_{01}\left(M-1\right) \\
B_{11}\left(0\right) & B_{11}\left(1\right) & B_{11}\left(2\right) & B_{11}\left(3\right) & \cdots & B_{11}\left(M-1\right) \\
B_{02}\left(0\right) & B_{02}\left(1\right) & B_{02}\left(2\right) & B_{02}\left(3\right) & \cdots & B_{02}\left(M-1\right) \\
\vdots & \vdots & \vdots & \vdots & \ddots & \vdots \\
B_{kk}\left(0\right) & B_{kk}\left(1\right) & B_{kk}\left(2\right) & B_{kk}\left(3\right) & \cdots & B_{kk}\left(M-1\right) \end{array} \right) \nonumber
\end{equation}
along with the vectors
\begin{equation}
a = \left( \begin{array}{c}
a_0 \\ a_1 \\ a_2 \\ a_3 \\ \vdots \\ a_{M-1} \end{array} \right) \qquad \mathrm{and} \qquad c = \left( \begin{array}{c}
C_{00} \\ C_{01} \\ C_{11} \\ C_{02} \\ \vdots \\ C_{kk} \end{array} \right) . \nonumber
\end{equation}

The matrix $B'$ is singular and often contains large, ill-conditioned submatrices. Because the rank of the matrix $B'$ is less than $M$, there exists no inverse and Eq. \ref{eq:ABCq} may have no solution. Therefore we approximate the Penrose-Moore pseudo-inverse via the singular value decomposition (SVD) method to find the least-norm solution for $a$ in the matrix equation given by Eq. \ref{eq:explicitBpac}.\cite{strang09.01} Note that there may be many other approaches to solve singular matrix equations with specific attributes based on the numerical values of the elements; however, our strategy is to use a generalized method that quickly approximates the polynomial coefficients so that we may incorporate this method into a Monte Carlo simulation as detailed in the discussion section.

\begin{figure}[th!]
\centering\includegraphics[scale=1]{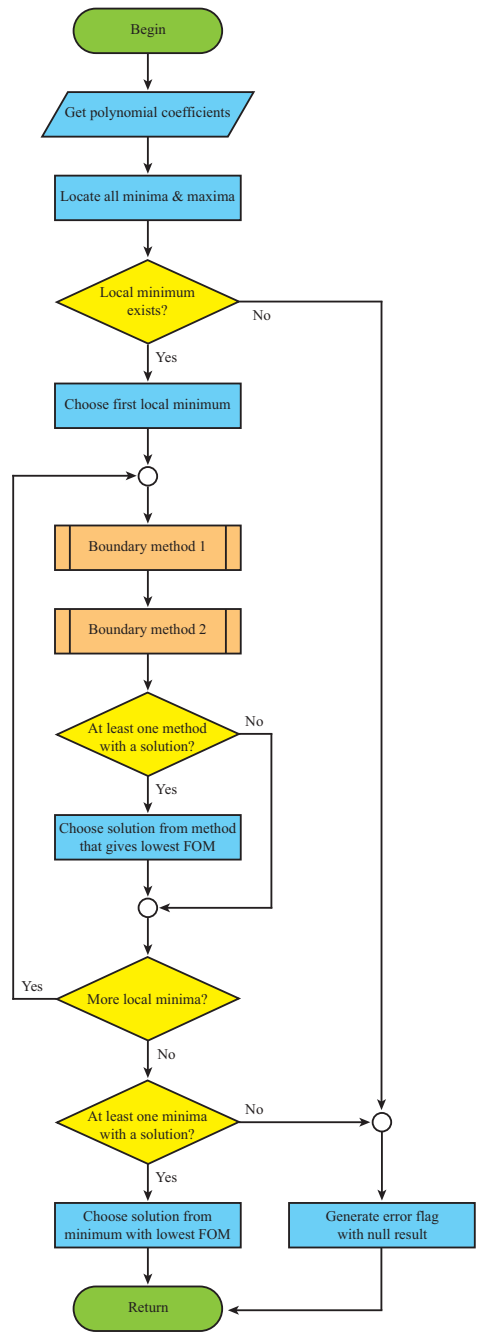}
\caption{A flow diagram for determining the region of space relevant for the finite input of transition moments and energies.}
\label{fig:logic1}
\end{figure}

We first rewrite the matrix $B'$ as the product of the unitary matrix of left-singular vectors, $U$, the diagonal matrix of singular values, ${\cal S}$, and the Hermitian adjoint of the unitary matrix of right-singular vectors, $V^\dag$,
\begin{equation}
B^\prime = U{\cal S}V^\dag .
\label{eq:Bsingular}
\end{equation}
The Penrose-Moore pseudo-inverse, which corresponds to the least-norm solution, follows as
\begin{equation}
B^{\prime +} = V{\cal S}^{+}U^\dag .
\label{eq:Bsingularinv}
\end{equation}
where the diagonal elements in ${\cal S}$ that are zero, or nearly zero, have their reciprocal set to exactly zero in ${\cal S}^{+}$. Note that Eq. \ref{eq:schrodME} can be displayed in a scale-free form \cite{shafe13.01} by substituting the quantities $e_n = E_n / E_{10}$, $x_\mathrm{max} = \hbar/\sqrt{2 m E_{10}}$, $\xi_{ij} = x_{ij} / x_\mathrm{max}$, and $b_q = a_q x_\mathrm{max} / E_{10}$; however, the numerical method used to determine the pseudo-inverse of the matrix $B$ is sensitive to the scale of the system. This approximation of the reciprocal diagonal elements of ${\cal S}$ as zero when the diagonal ${\cal S}$ elements are either near or below the specified tolerance, which depends on the machine's precision, can cause changes to the shape of the potential upon re-scaling. When the number of included states $N$ is large, which corresponds to potentials described by high-degree polynomials with $M$ coefficients (including the constant $a_0$), significant changes to the potential near the boundaries can occur with very small changes to the scale. A figure-of-merit (FOM) given by a scale-free sum-of-squares is used to measure the accuracy of the resulting potential based on the initially input and calculated output transition dipole moments and energies,
\begin{equation}
\mathrm{FOM} = \displaystyle \sum_{i=0}^{N} \sum_{j=0}^{N} \left(\displaystyle \frac{\left|\overline{x}_{ij}^\mathrm{calc}\right|^2}{\left|x_{\mathrm{max}}^\mathrm{calc}\right|^2} - \displaystyle \frac{\left|\overline{x}_{ij}^\mathrm{init}\right|^2}{\left|x_{\mathrm{max}}^\mathrm{init}\right|^2}\right)^2 .
\label{eq:FOM}
\end{equation}
The scale-free form of the FOM allows us to compare the transition dipole moments of the calculated potential to the initial values based on the shape of the potential. Because a polynomial potential is scale-invariant, we are only interested in shape of the calculated potential. Therefore, the transition dipole moments in relation to the ground state to first excited state transition dipole moment corresponds to a dimensionless FOM that compares the calculated potential's shape dependent parameters to those of the initial potential without penalizing potentials with similar shapes but slightly different sizes.

\begin{figure*}[th!]
\centering\includegraphics[scale=1]{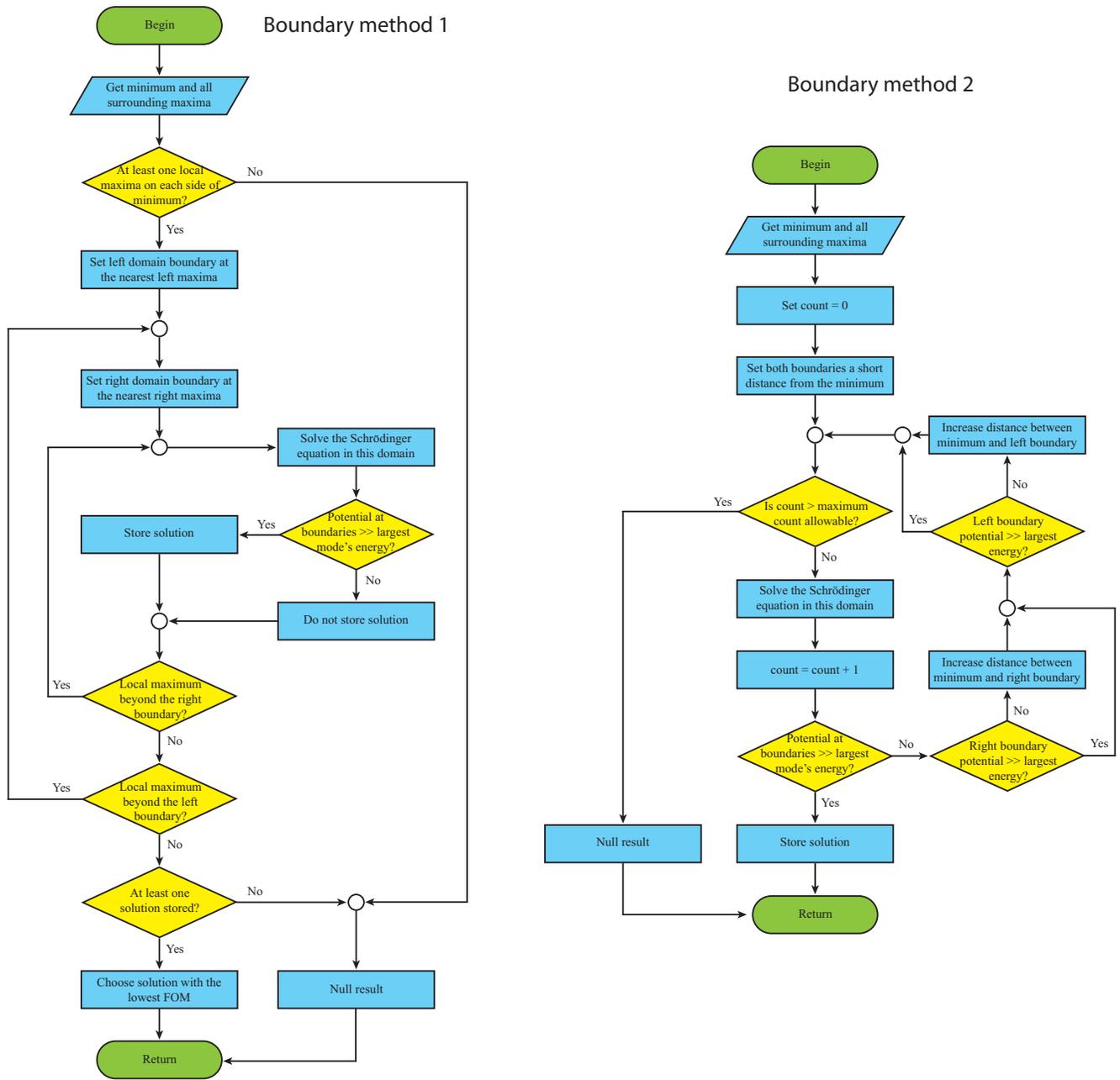}
\caption{Flow diagrams for determining the boundary locations using two different methods.}
\label{fig:logic2}
\end{figure*}

We can use a known potential energy function to check how well the procedure can reproduce this potential for a finite number of included states. The idea, then, is to (1) apply the above procedure to get the calculated transition dipole moments and energies from the input transition dipole moments and energies of the known potential; (2) determine the FOM using Equation \ref{eq:FOM}; and (3) scale the initial potential's width to give new transition dipole moments and energies, and repeat steps (1) and (2).  This procedure is repeated for a broad range of scaling factors to identify which one yields the lowest FOM. Note that changing the offset energy does not appear to significantly change the shape of the resulting potential function.

\begin{figure*}[th!]
\centering\includegraphics[scale=1]{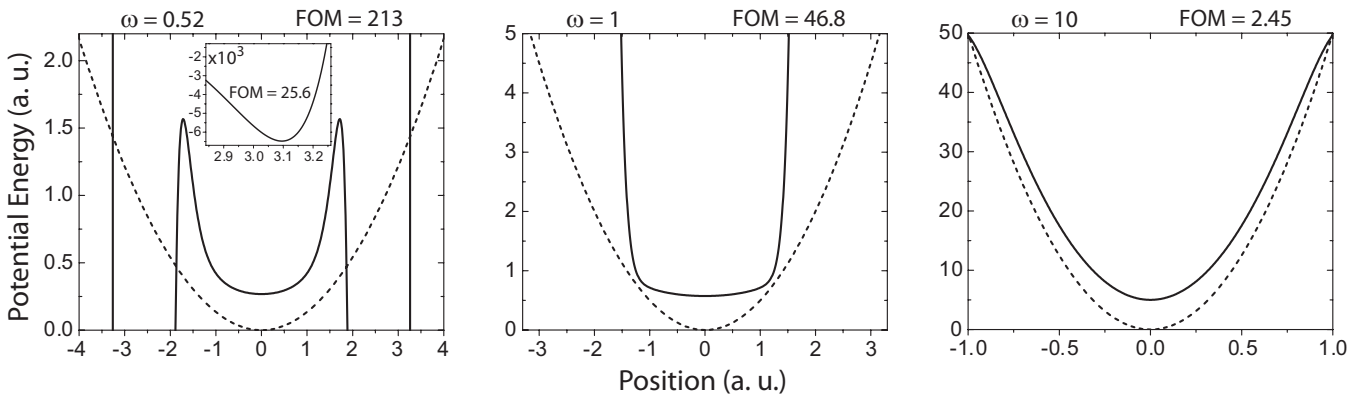}
\caption{The QHO potential (dashed) and the polynomial approximation to the potential (solid), which are calculated from the transition dipole moments and energies for different values of the angular frequency, $\omega$.  The inset shows a re-scaled view of the rightmost local minimum at $\sim 3.1\,$a. u. for the same graph when $\omega = 0.52\,$a. u.}
\label{fig:QHO6L}
\end{figure*}

The solution that we seek is the truncated power series that satisfies the given matrix equations. A high degree polynomial can have several local extrema. Finding the proper location of a potential well can be challenging as other minima may also be present outside, or inside, the domain of interest. In addition to the resultant truncation of a power series to a finite sum of powers used to describe the potential, the truncation of bound states limits the number of finite transition dipole moments used to find this potential. Thus, the information given to find the potential spans a finite domain, and the behavior of the potential beyond the relevant width must be removed. To find the appropriate domain, we first locate a minimum, which is characteristic of any potential well. Multiple minima are often a residual result of using high-degree polynomials and may exist outside of the domain. Multiple local minima, however, may also be present inside the domain, and choosing the proper domain can prove difficult with limited information from the finite number of transition dipole moments and energies. The method we used to identify the region of space relevant to the initial quantum system with only knowing the transition dipole moments and energies shown in Fig. \ref{fig:logic1}.

%The solution that we seek is the truncated power series that satisfies the given matrix equations. The number of transition dipole moments and energies is limited by the number of states included in the approximation. (I moved the first sentence to the beginning of the above paragraph, where this is redundant after discussing the problem of only fitting within a particular region in space where the well is located; this was starting to get redundant.)

When the highest power in the polynomial potential is odd with a nonzero coefficient or when the highest power is even with a negative coefficient, there must be a boundary introduced by us to neglect the other features of the potential outside the calculable domain (outside the potential well). Note that these set boundary locations can affect the calculated eigenenergies and wavefunctions. Thus, we consider potential wells with boundaries far from the walls and with potential energies at the boundaries that are much greater than the energy of the largest excited state. The boundary effects on the wavefunctions is minimal under these circumstances, where the tail of the probability distribution is approximately zero at these locations.
The methods we use to set the boundaries are both shown in Fig. \ref{fig:logic2}.

Note that we use Dirichlet boundary conditions, which remove the continuum state contributions.  We choose to set boundaries using these conditions because the least-norm solution only considers the behavior of the potential in the vicinity of the potential well.  The behavior outside of this region should be uncorrelated with the approximate solution to the potential in this region. An example would be an odd-degree polynomial solution with a non-zero value associated with the highest coefficient, which causes the potential functions to rapidly approach positive (negative) and negative (positive) infinity as the coordinate limits approach $\pm \infty$ (depending on the sign of the coefficient). Additionally, other unwanted features in the potential may manifest away from the bound state region, where the known transition dipole moments give little information about other regions.

For example, the eight-level model discussed in the next section has a resultant polynomial potential with a maximum power that is odd, and considering the potential out to $x = \pm \infty$ can result in an unphysical Hamiltonian if clear boundaries are not placed on the function. Likewise, polynomial potentials with a maximum power that is even can also occur, where an example would be the six-level model. Finding the best approximation to the potential from a finite matrix approach when the maximum power is even can result in a negative coefficient for the maximum power. Thus, although there is a good fit to the potential around the well, the fit to the potential far from the well can be quite poor. The reproduced potential when $\omega = 10\,$a. u. shown on the right side in Fig. \ref{fig:QHO6L} has a negative value for the largest coefficient of the polynomial, $a_{20} \approx -0.0801\,$a. u. This approximation to the harmonic oscillator corresponds to a good representation near the well minimum, but results in a barrier with a sunken peak if we do not set boundaries around the well. In the following section, the approximate potential is considered a good fit when the largest energy eigenvalue considered in the calculation is much less than the value of the potential at the Dirichlet boundary. Note that all values are given in atomic units, a. u., by setting $\hbar = 1$, $e = 1$, $m = 1$, and $\left(4\pi\epsilon_0\right)^{-1} = 1$.

\section{Discussion}

\subsection{Scale dependence}

The pseudo-inverse of the $B$ matrix only provides an approximation to the potential's power series coefficients when $B$ is singular. The SVD method gives the best approximation to the pseudo-inverse, where changes in the transition dipole moments and energies as parameters affect the results. This scale sensitivity is the primary reason we scanned the ground state to first excited state transition parameter over two orders-of-magnitude (correlating to changes in all transition energies and transition dipole moment parameters). Note that the optimal potentials did not occur at the same $E_{10}$ value, nor did there appear to be a common trend such as a convergence for decreasing or increasing the $E_{10}$ parameter. Those optimized potentials discovered by this method swept the entire range, where we only limited the range of possible values to allow for a reasonable computation time. Note that the shape of the resultant potential was negligibly sensitive to the choice of initial ground state energy, $E_{0}$.

We use the one-dimensional quantum harmonic oscillator (QHO), which has a well-known analytical solution, to evaluate the influence of scaling on the numerically approximated coefficients of the truncated power series that represents the potential with six states. Fig. \ref{fig:QHO6L} shows the harmonic potential and the numerically-determined potential using the actual transition dipole moments and energies as input parameters for differing length scales as quantified by the natural frequency of oscillation, $\omega$.  Also shown are the figures of merit. Very different potentials result.

\begin{figure}[t]
\centering\includegraphics[scale=1]{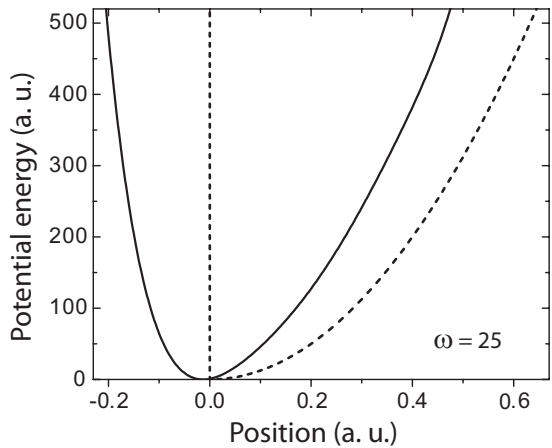}
\caption{The CQHO potential (dashed) and the polynomial approximation to the potential using a seven-state model (solid), which are computed using the actual transition dipole moments and energies as the input using the procedure described above for $\omega = 25\,$a. u.}
\label{fig:CQHO7L}
\end{figure}

The lowest angular frequency, shown in Fig.~\ref{fig:QHO6L}(a), has a harmonic-like shape near the local minimum at coordinate zero, but this local minimum is straddled by two narrow and deep wells, which affect the wavefunctions significantly.  The minimum is shown in the inset.  This corresponds to the least-norm solution about the region of interest near the origin's local minimum.\footnote{When the matrix equation $Ax = b$ has many solutions, the least-norm solution is the vector $x$ with the lowest norm.}  A large FOM is an indication that the polynomial approximation is poor for this scaling.

Re-scaling by increasing the energy spacing and decreasing the transition dipole moments by increasing $\omega$ removes the deep wells at the boundaries caused by the sharply varying polynomial.  The FOM also becomes smaller for this re-scaling.  The minimum flattens significantly around $\omega \approx 1$ a. u. as shown in Fig.~\ref{fig:QHO6L}(b). Fig. \ref{fig:QHO6L}(c) shows that increasing $\omega$ further gives a potential that best approximates the QHO which coincides with the smallest FOM.  The largest discrepancies between the true QHO potential and the polynomial approximation appear beyond the highest energy level used in the approximation, so meets the criteria of a good approximation.  %This scaling has the best FOM.

\subsection{Clipped quantum harmonic oscillator}

A nonzero hyperpolarizability requires the system to be asymmetric. The clipped quantum harmonic oscillator (CQHO) has known solutions that correspond to the odd-integer eigenvalues of the QHO. The Dirichlet boundary on the clipped side of the potential causes a significant asymmetric spatial profile. In addition to the large hyperpolarizability, the half potential boundary produces transition dipole moments that a truncated power series cannot represent well.

The actual transition dipole moments and energies for the CQHO are used to approximate the potential with a polynomial, and the results are shown in Fig. \ref{fig:CQHO7L} for a seven level model. The initial potential was scaled with $\omega = 25$ a.~u. The potential calculated from the transition dipole moments and energies is asymmetric, but the truncated power series is not a good approximation to the clipped side of the well.

Because the truncated power series is a poor approximation to the CQHO potential, the transition dipole moments and energies that result may be significantly different from those of the actual CQHO potential. In turn, the hyperpolarizability of the approximated potential may also be different than that calculated for the CQHO potential. As such, one approach is to increase the number of states (and increase the polynomial degree of the model potential) in an effort to increase the accuracy.

\begin{figure}[t]
\centering\includegraphics[scale=1]{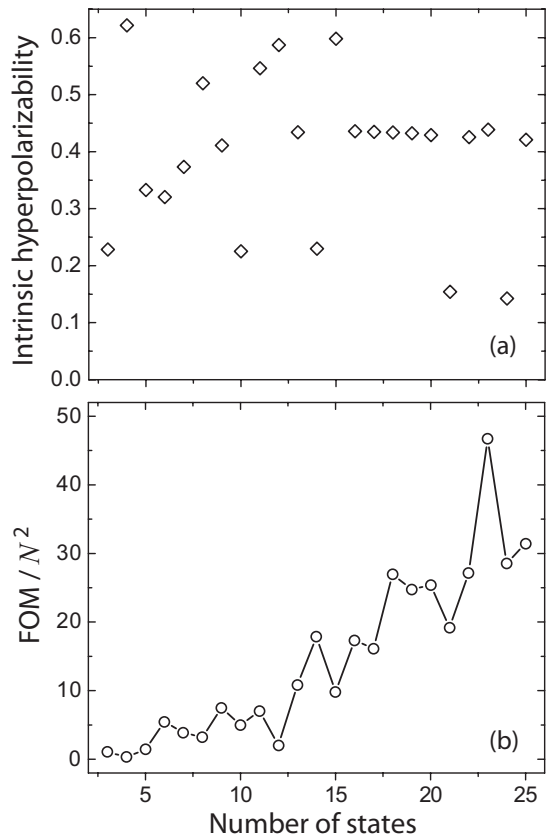}
\caption{(a) The intrinsic hyperpolarizability of a quantum clipped harmonic oscillator and (b) the FOM divided by the number of states squared as functions of the number of states used to calculate the intrinsic hyperpolarizability.}
\label{fig:betaFOM}
\end{figure}

The intrinsic hyperpolarizability as a function of the number of states used in the model is shown in Fig. \ref{fig:betaFOM}a.  The angular frequency was determined by a Monte Carlo technique, where $\omega = 10^{\eta}$ in atomic units with the random number $\eta$ ranging between 0.1 and 100. Seventy-five random samples were taken, and the one that gave polynomial potentials resulting in the smallest FOMs were used to calculate the hyperpolarizability.

The variations in the intrinsic hyperpolarizability as a function of the number of contributing states can be observed in Fig. \ref{fig:betaFOM}a.  They cluster near the accepted value of 0.57 for the CQHO.  Note that as the numbers of the states is increased, the term with the largest exponent in the polynomial increases, which increases the sensitivity of the calculation to the width of the CQHO potential.

The FOM$/N^2$ as a function of the number of states is given in Fig. \ref{fig:betaFOM}b. The FOM was divided by $N^2$ because the definition of the FOM is based on the sum-of-squares, where the number of terms in the summation is equal to number of states squared. Thus, the FOM is scale invariant for a specified number of states, but should not be used to compare systems with different numbers of states without this modification. Note that despite dividing by the total number of terms, the FOM still increases as the number of states increases. This increase of the FOM is due to the CQHO becoming wider with increasing energy, where the terms from higher state contributions are larger than those from lower states. Thus, the curve given in Fig. Fig. \ref{fig:betaFOM}b is expected.

\subsection{Determining potentials with large hyperpolarizabilities}

\begin{figure*}[t]
\centering\includegraphics[scale=1]{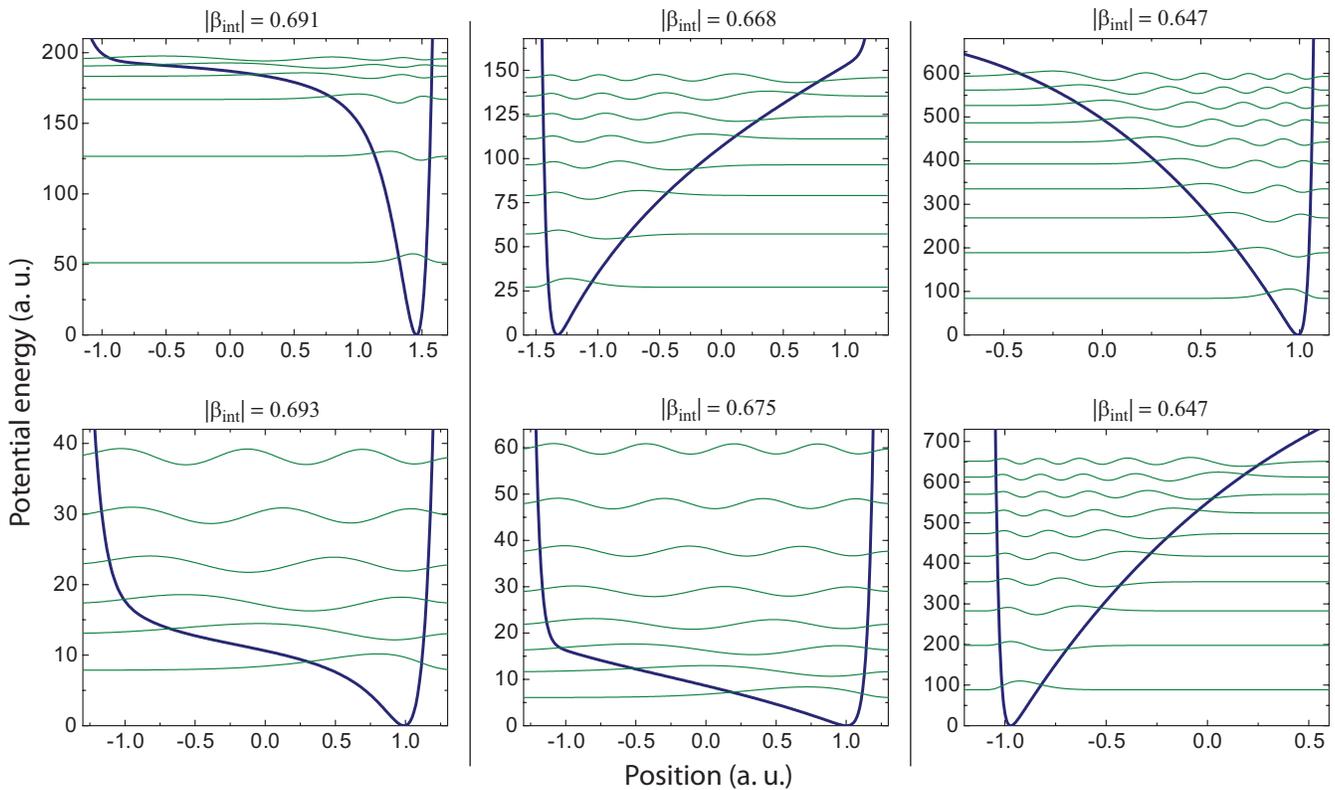}
\caption{Well-behaved potentials (thick blue line) with large calculated intrinsic hyperpolarizabilities using (left) six-, (middle) eight-, and (right) ten-levels. All wavefunctions (thin green lines) for the states included in the calculation are shown in each plot (with arbitrary scaled amplitudes), where the offsets correspond to the energy eigenvalues. Note that $a_0$ has been shifted so that the well's minimum is at zero.}
\label{fig:opt}
\end{figure*}

We seek to find potentials with large polarizabilities using only the required transition dipole moments and energies.  For a three-level model, the largest hyperpolarizability results when the second excited state approaches infinity.\cite{kuzyk00.01} Shafei and Kuzyk observed that when more states are included to calculate the fundamental limit, the limit gets larger than the value given by the three level ansatz in proportion to $\sqrt{N}$.\cite{shafe13.01} In this case, the maximum occurs when all excited states are degenerate except for one state, whose energy approaches infinity.  There are currently no known potentials that offer such a scheme, and our goal is to find one of these.  Since no system has ever been found that exceeds the fundamental limit calculated form the three-level ansatz, independent of how many states are included, it may be that such contrived systems do not result from any potential, and that exotic Hamiltonians that do not represent any real system may be required.

Including only three consecutive states results in a 5$^\mathrm{th}$-order polynomial, which does not reproduce the transition energy requirements, $E_{20} \gg E_{10}$, near the fundamental limit. A possible scheme to mimic such a high-energy excited state that is much greater than the preceding lower energy states would be to disallow transitions from the ground state to all states except for a transition to the first excited state and to the highest excited state, i.e. $\left|x_{0i}\right| \approx 0$ for $i = 2,...,N-1$.  These forbidden transitions to all other states make the system appear as a three-level model with transition energy $E_{N0}$, where $N$ is in essence the second excited state since no others contribute.  In this way, the transition energy of this contrived system can appear much greater than $E_{20}$ of the infinite square well, which has the greatest energy spacing of all simple potentials.

When setting all these transition dipole moments to zero, the TRK sum rules can be used to find the transition dipole moment $x_{0N}$ from the residual oscillator strength. When employing such a tactic, our technique generates potentials with large FOMs that often times were unbounded (resembling barriers). These unphysical results are attributed to setting these intermediate transitions to \textit{exactly} zero.

We found that assigning small random numbers to the transition dipole moments from the ground state to all states, with the exception of states 1 and N is a remedy of sorts, which swiftly reduces the number of unphysical potentials calculated.  Note that even small nonzero numbers for the intermediate transition dipole moments still produced large FOMs.  The polynomial approximation to the potentials that gave smaller FOMs required that the magnitudes of the randomized transition dipole moments be increased. Note that we also attempted to find potentials by setting the first two excited states to be near degenerate with transition dipole moments resembling those of a four-level model near the maximum and setting all transition dipole moments involving other states except the highest excited state to values near zero. This latter method proved less successful, although the complexity of many-state calculations did not allow us to completely exhaust the search.  %\textcolor{blue}{You may want to check out http://tinyurl.com/j3sb6ze.  This paper might elucidate the four-state model results better.  It was published in the journal Nonlinear Optics Quantum Optics.}

The well-behaved potentials with the largest intrinsic hyperpolarizabilies calculated from this method are shown in Fig. \ref{fig:opt} for six-, eight-, and ten-level models. These correspond to polynomials truncated at $x^{21}$, $x^{36}$, and $x^{55}$. Considering the degree of each polynomial the potentials shown in Fig. \ref{fig:opt} are relatively simple. Inside the boundaries of the walls, the potential function appears to monotonically increase from one end to the opposing side.  Thus, simple monotonic functions in deep wells produce the largest hyperpolarizability.

Recall that the calculations leading to Fig.~\ref{fig:opt} started by setting the transition dipole moments and energies to values that should lead to hyperpolarizabilities that break the limit calculated from the three level ansatz.  However, when solving the inverse problem with the polynomial method to determine the potential, the hyperpolarizabilities that are obtained do not exceed the limit, but are within 10\% of the largest calculated values.  Though this does not constitute proof, these observations are consistent with the assertion that the three-level ansatz is a requirement for calculating the limits. Using more states leads to a higher limit, which for an infinite number of states becomes unbounded.  Given that such divergent behavior has not been observed suggests that these types of systems are not of the type represented by a Schr\"{o}dinger equation based on a mechanical Hamiltonian. Thus, the TRK sum rules allows for a larger nonlinear response, but the Hamiltonians that are required would need to be of an exotic nature that may not correspond to real systems.

\subsection{Half potentials of positive power}

The power series potentials in Figure \ref{fig:opt}, which are endowed with some of the largest intrinsic hyperpolarizabilities, resemble simple half potentials.  To test the hypothesis that power law half potentials may yield the largest hyperpolarizabilities, we test potentials of the form
\begin{equation}
V\left(x\right) = \begin{dcases} x^\eta & \mathrm{for} \quad x > 0 \\
\infty & \mathrm{for} \quad x \leq 0 . \end{dcases}
\label{eq:powpot}
\end{equation}
The sharp Dirichlet boundary at $x = 0$ of the half potential makes it asymmetric for finite powers.  Negative powers are not considered, where these potentials are expected to yield lower hyperpolarizabilities due to the continuum state contribution.

\begin{figure}[t]
\centering\includegraphics[scale=1]{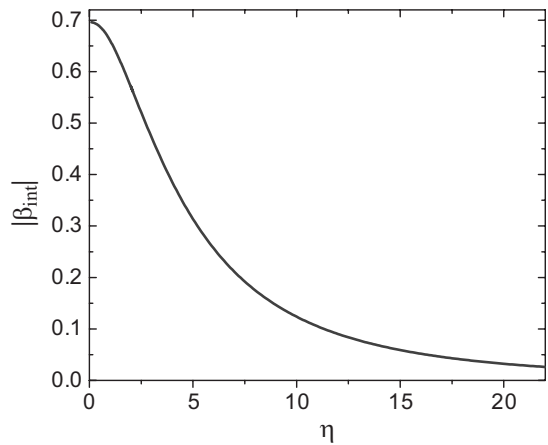}
\caption{The intrinsic hyperpolarizability as a function of the power for half potentials of simple power functions given by Eq. \ref{eq:powpot}.}
\label{fig:power}
\end{figure}

The calculated intrinsic hyperpolarizability as a function of $\eta$ is shown in Fig. \ref{fig:power}. The curve is a near fit to a Lorentzian centered at the origin with an approximate full width of 9. In the limit of infinite exponent, the potential approached a centro-symmetric box-like potential well, where the hyperpolarizability vanishes. The intrinsic hyperpolarizability approaches a maximum of approximately 0.696 as the exponent approaches zero, where the potential is highly asymmetric. Using just a single parameter power function, we observe an intrinsic hyperpolarizability near the largest value previously reported via optimization techniques, which are of the order of $\sim 0.709$.\cite{zhou06.01,zhou07.02,ather12.01}

\section{Conclusion}

We have developed a method to determine an approximate polynomial potential from transition dipole moments and energies.  This method requires the use of a pseudo-inverse to approximate the potential due to the singular/ill-conditioned nature of the finite matrix involved in the calculation of coefficients for a truncated power series.  We have also created an algorithm that locates minima and places boundaries around the potential.  These boundaries are necessary due to the finite range of validity caused by the truncation of eigenstates.

We observe that the range of transition moments and energies that give hyperpolarizabilities near the limit do not produce well-behaved polynomial potentials when solving the inverse problem.  This deviation is characterized by large FOMs as well as polynomial potentials with much smaller intrinsic hyperpolarizabilities than those determined from the transition moments and energy spectrum that come from the TRK sum rules.  In such cases, we often find resultant potentials with no minimum anywhere in space (barriers).  This is found to be the case for the transition dipole moments and energies in the three-state model where the intrinsic hyperpolarizability is at the limit.  Those initial large-hyperpolarizability cases that give resultant potentials with minima yield large FOMs, which indicate that the potential determined by solving the inverse problem is a poor representation of the potential.

In contrast, transition dipole moments and energies that give a nonlinear response below the apparent limit yield a good approximation to the potential with our method.  As the hyperpolarizability reaches the limit, the procedure yields values that are highly unreliable, and those that exceed the limit through the many-state catastrophe yield potentials that give hyperpolarizabilities below the limits when our method gives a reliable potential well as determined from the FOM.  The fact that the procedure breaks down when the transition dipole moments and energies give a nonlinear-optical response approaching the limit may be an indicator that the limit is unattainable for a system that is described by a polynomial potential or perhaps for any physical potential.  This might imply that reaching the limit requires exotic Hamiltonians that go beyond the mechanical Hamiltonian on which the Schr\"{o}dinger equation is based.  Furthermore, this work provides insights into the three-level ansatz, which appears to be a requirement for not overestimating the fundamental limit, an assertion that might be unprovable.\cite{shafe13.01}

\begin{figure}[t]
\centering\includegraphics[scale=1]{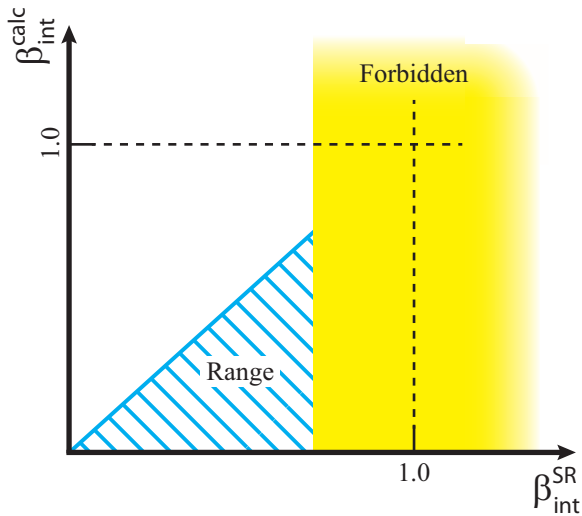}
\caption{Highly schematic representation of the hyperpolarizability range calculated from the inverse method, $\beta_{int}^{calc}$, as a function of the hyperpolarizability calculated from transition moments and energies constrained by the sum rules, $\beta_{int}^{SR}$.}
\label{fig:Summary}
\end{figure}

Figure \ref{fig:Summary} is a highly schematic summary of the intrinsic hyperpolarizability range calculated from the inverse method, $\beta_{int}^{calc}$, as a function of the intrinsic hyperpolarizability calculated from transition moments and energies constrained by the sum rules, $\beta_{int}^{SR}$.  When the sum-rule-constrained transition moments and energies yield an intrinsic hyperpolarizability below about 0.7 ($\beta_{\text{int}}^{\text{SR}} \lesssim 0.7 $), the procedure described here gives a potential energy function that when used in the Schrodinger Equation leads to an intrinsic hyperpolarizability that is smaller than the one obtained with the sum rules $\beta_{\text{int}}^{\text{calc}} \lesssim \beta_{\text{int}}^{\text{SR}}  $), as represented by the blue-hashed region.  When the sum-rule-constrained energy and transition moments are chosen to be above 0.7 ($\beta_{\text{int}}^{\text{SR}} \gtrsim 0.7$) our procedure breaks down and does not produce a potential energy function.  This region of unstable inverse, shaded in yellow, indicates that our method cannot find a potential with an intrinsic hyperpolarizability near the limit, so may be an indication of the impossibility of reaching the limit.  This is yet another method that implies that the true limit may be on the order of 0.709 of the intrinsic value.

We have discovered potentials from this method that approach the apparent limits of the intrinsic hyperpolarizability.  Attempts to find larger intrinsic hyperpolarizabilities via this method resulted in strange potentials that possessed significantly lower intrinsic values. The potentials that we discovered with the largest intrinsic hyperpolarizabilities resembled simple monotonically increasing half potentials.  This motivated us to evaluate power law potentials, with positive exponent, which are found to fall just short of the apparent limit of 0.709 when the power vanishes.  It is interesting that varying just one parameter yields an intrinsic hyperpolarizability that nearly spans the full range that is possible.

Though the approach that we developed here is a useful tool for determining a potential energy function from transition dipole moments and energy spectrum, it is telling us that regimes where this technique breaks down are those that obey the TRK sum rules but not the three-level ansatz.  As such, this type of approach may be a more powerful tool in investigations of the structure of the theory of limits for systems at or beyond the limit.  For this approach to yield a rigorous test of hypotheses about the behavior of the nonlinear response at the limit requires further refinements, which we are presently pursuing.

\section*{Acknowledgments}

MGK acknowledges the National Science Foundation (ECCS-1128076) and the Meyer Distinguished Professorship in the Sciences for generously supporting this work.

%\bibliographystyle{osajnl}
%\bibliography{\bibs}

\begin{thebibliography}{10}
\newcommand{\enquote}[1]{``#1''}
\expandafter\ifx\csname url\endcsname\relax
  \def\url#1{\texttt{#1}}\fi
\expandafter\ifx\csname urlprefix\endcsname\relax\def\urlprefix{URL }\fi
\providecommand{\eprint}[2][]{\url{#2}}

\bibitem{unknown15.oct}
\enquote{{Global photonics components business},}
  SPIE Professional (October, 2015). doi: 10.1117/2.4201510.03.

\bibitem{kuzyk00.01}
M.~G. Kuzyk, \enquote{{Physical Limits on Electronic Nonlinear Molecular
  Susceptibilities},} Phys. Rev. Lett. \textbf{85}, 1218 (2000).

\bibitem{thoma25.01}
W.~Thomas, \enquote{Uber die Zahl der Dispersionselektronen, die einem
  station\"{a}ren Zustande zugeordnet sind.(Vorlaufige Mitteilung),}
  Naturwissenschaften \textbf{13}(28), 627--627 (1925).

\bibitem{reich25.01}
F.~Reiche and W.~Thomas, \enquote{{\"{U}ber die Zahl der Dispersionselektronen,
  die einem station\"{a}ren Zustand zugeordnet sind},} Z. Phys. \textbf{34}(1),
  510--525 (1925).

\bibitem{kuhn25.01}
W.~Kuhn, \enquote{{\"{U}ber die Gesamtst\"{a}rke der von einem Zustande ausgehenden
  Absorptionslinien},} Z. Phys. \textbf{33}(1), 408--412 (1925).

\bibitem{ather12.01}
T.~J. Atherton, J.~Lesnefsky, G.~A. Wiggers, and R.~G. Petschek,
  \enquote{{Maximizing the hyperpolarizability poorly determines the
  potential},} J. Opt. Soc. Am. B \textbf{29}(3), 513--520 (2012).

\bibitem{zhou06.01}
J.~Zhou, M.~G. Kuzyk, and D.~S. Watkins, \enquote{{Pushing the
  hyperpolarizability to the limit},} Opt. Lett. \textbf{31}, 2891 (2006).

\bibitem{zhou07.02}
J.~Zhou, U.~B. Szafruga, D.~S. Watkins, and M.~G. Kuzyk, \enquote{{Optimizing
  potential energy functions for maximal intrinsic hyperpolarizability},} Phys.
  Rev. A \textbf{76}, 053831 (2007).

\bibitem{shafe12.01}
S.~Shafei, R.~Lytel, and M.~G. Kuzyk, \enquote{{Geometry-controlled nonlinear
  optical response of quantum graphs},} J. Opt. Soc. Am. \textbf{29}(12),
  3419--3428 (2012).

\bibitem{lytel15.01}
R.~Lytel, S.~M. Mossman, and M.~G. Kuzyk, \enquote{{Optimization of eigenstates
  and spectra for quasi-linear nonlinear optical systems},} J. Nonlinear Opt.
  Phys. Mat. \textbf{24}(2), 1550018 (2015).

\bibitem{lytel15.02}
R.~Lytel, S.~M. Mossman, and M.~G. Kuzyk, \enquote{{Phase disruption as a new
  design paradigm for optimizing the nonlinear-optical response},} Opt. Lett.
  \textbf{40}(20), 4735--4738 (2015).

\bibitem{kuzyk13.01}
M.~G. Kuzyk, J.~P\'{e}rez-Moreno, and S.~Shafei, \enquote{{Sum rules and
  scaling in nonlinear optics},} Phys. Rep. \textbf{529}(4), 297--398 (2013).

\bibitem{Tripa04.01}
K.~Tripathy, P.~Moreno, M.~G. Kuzyk, B.~J. Coe, K.~Clays, and A.~M. Kelley,
  \enquote{{Why hyperpolarizabilities Fall Short of the Fundamental Quantum
  Limits},} J. Chem. Phys. \textbf{121}(16), 7932--7945 (2004).

\bibitem{perez07.01}
J.~P\'{e}rez-Moreno, Y.~Zhao, K.~Clays, and M.~G. Kuzyk, \enquote{{Modulated
  conjugation as a means for attaining a record high intrinsic
  hyperpolarizability},} Opt. Lett. \textbf{32}(1), 59--61 (2007).

\bibitem{perez07.02}
J.~P\'{e}rez-Moreno, I.~Asselberghs, Y.~Zhao, K.~Song, H.~Nakanishi, S.~Okada,
  K.~Nogi, O.-K. Kim, J.~Je, J.~Matrai, M.~De~Mayer, and M.~G. Kuzyk,
  \enquote{{Combined molecular and supramolecular bottom-up nano-engineering
  for enhanced nonlinear optical response: Experiments, modelling and
  approaching the fundamental limit},} J. Chem. Phys. \textbf{126}(7), 074705
  (2007).

\bibitem{cole02.01}
J.~M. Cole, R.~C.~B. Copley, G.~J. McIntyre, J.~A.~K. Howard, M.~Szablewski,
  and G.~H. Cross, \enquote{{Charge-density study of the nonlinear optical
  precursor DED-TCNQ at 20 K},} Phys. Rev. B \textbf{65}(12), 125107 (2002).

\bibitem{cole03.01}
J.~M. Cole, \enquote{{Organic materials for second-harmonic generation:
  advances in relating structure to function},} Phil. Trans. R. Soc. Lond.
  \textbf{361}(1813), 2751--2770 (2003).

\bibitem{higgi12.01}
A.~P. Higginbotham, J.~M. Cole, M.~A. Blood-Forsythe, and D.~D. Hickstein,
  \enquote{{Identifying and evaluating organic nonlinear optical materials via
  molecular moments},} J. Appl. Phys. \textbf{111}(3), 033512 (2012).

\bibitem{May07.01}
J.~C. May, I.~Biaggio, F.~Bures, and F.~Diederich, \enquote{{Extended
  conjugation and donor-acceptor substitution to improve the third-order
  optical nonlinearity of small molecules},} App. Phys. Lett. \textbf{90},
  251106 (2007).

\bibitem{stefk13.01}
M.~Stefko, M.~D. Tzirakis, B.~Breiten, M.-O. Ebert, O.~Dumele, W.~B. Schweizer,
  J.-P. Gisselbrecht, C.~Boudon, M.~T. Beels, I.~Biaggio, and F.~Diederich,
  \enquote{{Donor-Acceptor (D-A)-Substituted Polyyne Chromophores: Modulation
  of Their Optoelectronic Properties by Varying the Length of the Acetylene
  Spacer},} Chem. Eur. J. \textbf{19}(38), 12693--12704 (2013).

\bibitem{Kang05.01}
H.~Kang, A.~Facchetti, P.~Zhu, H.~Jiang, Y.~Yang, E.~Cariati, S.~Righetto,
  R.~Ugo, C.~Zuccaccia, A.~Macchioni, C.~L. Stern, Z.~Liu, S.~T. Ho, and T.~J.
  Marks, \enquote{{Exceptional Molecular Hyperpolarizabilities in Twisted
  $\pi$-Electron System Chromophores},} Angew. Chem. Int. Ed. \textbf{44}
  (2005).

\bibitem{Kang07.01}
H.~Kang, A.~Facchetti, H.~Jiang, E.~Cariati, S.~Righetto, R.~Ugo, C.~Zuccaccia,
  A.~Macchioni, C.~L. Stern, Z.~F. Liu, S.~T. Ho, E.~C. Brown, M.~A. Ratner,
  and T.~J. Marks, \enquote{{Ultralarge hyperpolarizability twisted
  $\pi$-electron system electro-optic chromophores: Synthesis, solid-state and
  solution-phase structural characteristics, electronic structures, linear and
  nonlinear optical properties, and computational studies},} J. Am. Chem. Soc.
  \textbf{129}(11), 3267--3286 (2007).

\bibitem{brown08.01}
E.~C. Brown, T.~J. Marks, and M.~A. Ratner, \enquote{{Nonlinear Response
  Properties of Ultralarge Hyperpolarizability Twisted p-System Donor-Acceptor
  Chromophores. Dramatic Environmental Effects on Response},} J. Phys. Chem. B
  \textbf{112}(1), 44--50 (2008).

\bibitem{shi15.02}
Y.~Shi, A.~J.-T. Lou, G.~S. He, A.~Baev, M.~T. Swihart, P.~N. Prasad, and T.~J.
  Marks, \enquote{{Cooperative Coupling of Cyanine and Tictoid Twisted
  p-Systems to Amplify Organic Chromophore Cubic Nonlinearities},} J. Am. Chem.
  Soc. \textbf{137}(14), 4622--4625 (2015).

\bibitem{knopp12.01}
S.~Knoppe, A.~Dass, and T.~B\"{u}rgi, \enquote{{Strong non-linear effects in
  the chiroptical properties of the ligand-exchanged Au$_{38}$ and Au$_{40}$
  clusters},} Nanoscale \textbf{4}(14), 4211--4216 (2012).

\bibitem{knopp15.01}
S.~Knoppe, M.~Vanbel, S.~v. Cleuvenbergen, L.~Vanpraet, T.~B\"{u}rgi, and
  T.~Verbiest, \enquote{{Nonlinear Optical Properties of Thiolate-Protected
  Gold Clusters},} J. Phys. Chem. C \textbf{119}(11), 6221--6226 (2015).

\bibitem{watki12.01}
D.~S. Watkins and M.~G. Kuzyk, \enquote{{Universal properties of the optimized
  off-resonant intrinsic second hyperpolarizability},} J. Opt. Soc. Am. B
  \textbf{29}(7), 1661--1671 (2012).

\bibitem{shafe13.01}
S.~Shafei and M.~G. Kuzyk, \enquote{{Paradox of the many-state catastrophe of
  fundamental limits and the three-state conjecture},} Phys. Rev. A
  \textbf{88}(2), 023863 (2013).

\bibitem{burke13.01}
C.~J. Burke, T.~J. Atherton, J.~Lesnefsky, and R.~G. Petschek,
  \enquote{{Optimizing the second hyperpolarizability with minimally
  parametrized potentials},} J. Opt. Soc. Am. B \textbf{30}(6), 1438--1445
  (2013).

\bibitem{dawson15.01}
N.~J. Dawson, \enquote{{Lowest-order relativistic corrections to the
  fundamental limits of nonlinear-optical coefficients},} Phys. Rev. A
  \textbf{91}(1), 013832 (2015).

\bibitem{kac66.01}
M.~Kac, \enquote{{Can One Hear the Shape of a Drum?}} Am. Math. Monthly
  \textbf{73}(4), 1--23 (1966).

\bibitem{kuzyk08.01}
M.~C. Kuzyk and M.~G. Kuzyk, \enquote{{Monte Carlo Studies of the Fundamental
  Limits of the Intrinsic Hyperpolarizability},} J. Opt. Soc. Am. B.
  \textbf{25}(1), 103--110 (2008).

\bibitem{strang09.01}
G.~Strang, \emph{Introduction to Linear Algebra}, 4th ed. (Wellesley-Cambridge
  Press, Wellesley, 2009).

\bibitem{Note1}
When the matrix equation $Ax = b$ has many solutions, the least-norm solution
  is the vector $x$ with the lowest norm.

\end{thebibliography}

%Manual citation list
%\begin{thebibliography}{1}
%\bibitem{Zhang:14}
%Y.~Zhang, S.~Qiao, L.~Sun, Q.~W. Shi, W.~Huang, %L.~Li, and Z.~Yang,
 % \enquote{Photoinduced active terahertz metamaterials with nanostructured
  %vanadium dioxide film deposited by sol-gel method,} Opt. Express \textbf{22},
  %11070--11078 (2014).
%\end{thebibliography}

% Please include bios and photos of all authors for aop articles
%\ifthenelse{\equal{\journalref}{aop}}{%
%\section*{Author Biographies}
%\begingroup
%\setlength\intextsep{0pt}
%\begin{minipage}[t][6.3cm][t]{1.0\textwidth} % Adjust height [6.3cm] as required for separation of bio photos.
%  \begin{wrapfigure}{L}{0.25\textwidth}
%    \includegraphics[width=0.25\textwidth]{NJD1.eps}
%  \end{wrapfigure}
%  \noindent
%  {\bfseries Nathan J. Dawson} received his PhD (Physics) in 2010 from Washington State University. He is an assistant professor in the department of physics at The College of The Bahamas. He also holds the position of adjunct assistant professor in the department of physics and astronomy at Washington State University. His primary research interests include organic nonlinear optics and polymer lasers.
%\end{minipage}
%\endgroup
%}{}

\end{document}